# Mastering Music Instruments through Technology in Solo Learning Sessions

**Karola Marky**
TU Darmstadt
Darmstadt, Germany
marky@tk.tu-darmstadt.de

**Julien Gedeon**
TU Darmstadt
Darmstadt, Germany
gedeon@tk.tu-darmstadt.de

**Andreas Weiß**
Musikschule Schallkultur
Kaiserslautern, Germany
andreas.weiss@musikschule-schallkultur.de

**Sebastian Günther**
TU Darmstadt
Darmstadt, Germany
guenther@tk.tu-darmstadt.de

## Abstract
Mastering a musical instrument requires time-consuming practice even if students are guided by an expert. In the overwhelming majority of the time, the students practice by themselves and traditional teaching materials, such as videos or textbooks, lack interaction and guidance possibilities. Adequate feedback, however, is highly important to prevent the acquirement of wrong motions and to avoid potential health problems. In this paper, we envision musical instruments as smart objects to enhance solo learning sessions. We give an overview of existing approaches and setups and discuss them. Finally, we conclude with recommendations for designing smart and augmented musical instruments for learning purposes.

## ACM Classification Keywords
H.5.m [Information interfaces and presentation (e.g., HCI)]: Miscellaneous

## Introduction
Learning a musical instrument is a time-consuming task that requires a lot of practice. Even if students are guided by an experienced musician or expert, the students prac-tice by themselves in the overwhelming majority of the time since experienced musicians or experts are limited resources for several reasons. Those include availability and the cost of lessons. Traditional self-teaching materials, such as videos or textbooks, lack interaction and guidance possibilities. Thus, during solo practice sessions, the students do not receive immediate feedback. Adequate feedback, however, is highly important because the students might acquire wrong movements or postures that are difficult and time-consuming to correct later on. Missing feedback might even lead to health problems such as repetitive strain injury [18] or sore arms. Hence, students – especially beginners – require assistance during solo learning sessions that can provide them with adequate and immediate feedback.

More and more daily devices are turned into smart objects by either equipping them with sensors, actuators, and further computation and connection capabilities [13]. This technology can be leveraged to provide students with adequate feedback during solo learning sessions [9, 12] to enhance their music playing abilities. In this paper, we give an overview of existing technologies that augment and transform musical instruments into smart objects and derive recommendations for an optimal learning experience and accessibility during solo practice sessions.

## Augmented and Smart Musical Instruments

In general, there are two possibilities to turn a musical instrument into a smart object: (1) augmentation by external devices and (2) integration of sensors and actuators. If the musical instrument is augmented by an external device, a conventional musical instrument, such as a guitar or a piano, it remains unaltered and the assistance comes from external devices. This could, for instance, be a screen [14], a projector [12, 15] or actuators [9]. If the musical instrument includes sensors and actuators, either a conventional musical instrument has to be transformed, or the musical instrument has to be built specifically to integrate the sensors and actuators. In the following, we give an overview of existing setups and discuss their benefits and drawbacks.

*Screen-Based Augmentation*
The simplest setup for screen-based augmentation is adding a screen that displays information. *Synthesia* [21] is a setup in which a screen is placed on top of a piano. The screen displays bars that are moving from the screen top towards the keys of a virtual piano. The length of the bar depicts the length of a tone, and the colors of the bars correspond to the hands. Numbers from 1 to 5 depict the finger that should press the respective key. While this visualization is widely adopted among autodidacts, a cognitive mapping from the keys on the screen to the real keys and fingers is required. This constitutes a difficulty because the student has to translate the depicted finger position that does not match his or her view. Information on the correct finger and body posture is not provided. The visual representation of tones as bars is more intuitive than sheet notation and can, therefore, be used more quickly [17].

Cakmakci *et al.* [2] use optical markers to detect finger positions on a bass guitar and overlay a camera image with target finger positions as well as directions. Liarokapis [11] introduces markers that can be placed in the student's environment. If the camera recognizes a marker, it augments it with finger placing information, e.g., a schematic depiction of a guitar chord. Thus, the student is not limited to look at one specific spot. The position tracking in these setups is limited to the position of the fingertips, but feedback regarding the finger posture, such as the angle of the joints or the required pressure, is not provided. Motogawa *et al.* [14] extended the setups mentioned above by adding a 3D-model of the ideal hand posture to the image. This introduces feedback regarding the finger posture and the students can check if their hand postures correspond to the presented one.

All setups presented up to now support the students in placing their fingers by showing them finger targets. Kerdvibulvech and Saito [10] display the currently played chord as well as target finger positions on a screen. Therefore, they track the guitar neck and the fingers of the students with markers. Because the students have to look at a screen, their view of the finger targets is inverted. Thus, similarly to traditional depictions on paper, a constant perceptual mapping between the screen and the real world is required.

*Light-Based Augmentation*
In light-based augmentation, visual light cues are used. There are three possible light sources: (1) an integrated light source, (2) a mounted light source and (3) projected light. Integrated light sources can, for instance, be built into the fretboard of a (bass) guitar [8, 20] or the key of a piano or keyboard [7]. A plethora of commercially available light-based products is already on the market, such as the *gTar* [8] or *FretLight* [20]. Mounted light sources can be mounted on conventional musical instruments [3]. The finger targets are visible and displayed at a position where no perceptual mapping is required.

In projection-based augmentation, the information is projected either directly on the surface of the instrument, or a projection surface close to the instrument. Takegawa and Terada [22] use a projector to support the learning of sheet music notation by projecting visual connections between written notes and piano keys.

The setup *guitAR* [12] projects finger targets on a guitar fretboard to show the students where to press. In the approach by Yang and Essl [24], *a Game of Tones* [15] and *P.I.A.N.O.* [17] a projection surface above a piano is used to project the next tones above the respective key. Both setups can be seen as an extension of *Synthesia* [21] to a projection surface. The projection surface, however, is not part of the piano and needs to be placed on top of it which restricts this technology to certain piano types. Furthermore, feedback regarding the finger and hand postures are not given in all setups.

In general, light-based setups cannot provide feedback regarding postures, since the students only know where to press but not how. The number of visible finger positions at a given time is limited because the student might not realize which key or string to press. The setups cannot react to the student's individual speed, since they can only display a sequence of chords or tones in a pre-defined tempo without adapting to the student's playing tempo.

*Actuation- and Sensor-Based Augmentation*
Augmentation by actuation means that a component sets the students' bodies into motion. The students' postures or movements can be captured by sensors. Similarly to light-based augmentation, the actuators and sensors can either be integrated into the musical instruments or can be mounted on them.

*MusicJacket* is a jacket that aims to teach body posture and the bowing techniques to violin players via vibrotactile feedback [23]. The students' motions are captured by several sensors while their movements are corrected by seven vibration motors on the upper body.

*Mobile Music Touch (MTT)* is a haptic glove that aims to teach the fingering for playing the piano [6]. While the student listens to a song, a vibration indicates the finger that is used to press this note.

The system *EMGuitar* uses electromyography to assess the student's finger postures on the guitar [9]. Electromyography captures the electrical activity that is produced by a muscle. Therefore, the student wears an armband-like construction that captures the muscle activity. Based on that it adjusts the tempo of displayed chords such that the student can play without interruptions.

Shin *et al.* propose a guitar setup that combines integrated LEDs, piezo sensors and microcontrollers [19]. They use different light colors for the different fingers and an application to provide feedback. The feedback is limited to the duration of a chord or tone and the position of the finger on the fretboard.

### Recommendations

From the related work presented above and an expert discussion, we derive recommendations for designing and building smart musical instruments that are based on strings or keyboards:

1. Place guidance cues in the student's natural field of view.

2. Reduce or avoid perceptual mappings by placing the (visual) cues as close to the targets as possible, ideally directly on them. Furthermore, avoid inverted images.

3. Include postures, e.g. of the fingers or the body, in the cues.

4. Avoid additional (mounted) components, such as markers, because they might disturb or distract the students.

5. Wearable devices should not interfere with playing the musical instrument.

6. Be reactive on demand to adapt to the students' needs and the current learning curve (such as tempo).

7. Projection-based guidance and feedback should not require an additional (mounted) projection surface.

### Discussion

New students are confronted with a rather huge cognitive load: they have to learn the setup of the musical instrument, instrument-specific music notation, postures of different body parts, and they have to develop a feeling for rhythm and tones. From a student's perspective, it makes sense to reduce this cognitive load as much as possible and gradually introduce one aspect after another instead of all at once. Smart and augmented musical instruments aim to support students by reducing the cognitive load in several ways:

- alternative and more intuitive visualizations of sheet music (cf. [21, 17])

- visual or haptic cues for finger and body placement (cf. [17])

- feedback regarding (in-)correct finger placement and body posture (cf. [23])

While those techniques are beneficial in the learning scenario, their long-term effects are not well understood yet. It remains unclear whether students might develop a dependency on those technologies, meaning they might overly rely on them. The purpose of the learning setup must be communicated clearly to the students which might result in overhead since using the setup requires introduction time and further practice.

The presented setups support students in different tasks but are mainly focused on finger positioning, i.e., they teach where to press a string or key but not how to do it. In this scope, direct approaches that do not require perceptual mappings are beneficial. The visual cues are at the real finger target and the students directly recognize whether they touch the correct spots. Wearing a head-mounted display that overlays the real world with the augmented information could be an appropriate solution for the target display. Since optical markers on the instrument and the fingertips might influence the student's movements, the finger position could be estimated by approaches that are based on depth-images [16].

Even if the students touch the correct spots, the finger, arm or body posture might be incorrect. Joints, especially those of the fingers, could be overstretched and the motions might be too cramped. Over a long period this might lead to persisting health problems and might even slow down further progress (in more complicated compositions).

Up to this point, we presented the means to enhance the learning of traditional musical instruments. But the advance of technologies enables a new generation of musical instruments. Musical instruments have already made several progressions such as the one from the harpsichord to the piano or from the baroque cello to the modern version of the cello. Screens, lights, and actuators bear the potential to transform existing musical instruments into new instruments with new features. For instance, the hand position problems that are present in nearly all musical instruments could be mitigated by sensor technology. This could, for instance, result in a piano with an ergonomic keyboard. Furthermore, learning-disabled musicians might benefit from simplified smart instrument setups as shown by Harrison *et al.* [5].

While our recommendations focus on solo learning sessions, augmented and smart musical instruments also enable new forms of remote collaboration. While approaches for remote music performances are already available in the literature [4, 1], remote music lessons nowadays rely on simple setups, such as video calling technology, that inherit several drawbacks from video learning materials although an experienced musician or expert is present and available. Sensors, lights, and actuators that are controlled by the music teacher might offer an enhanced learning experience. We envision that the teachers could, for instance, show a combination of tones on their musical instruments or play together with the students. The body and or finger postures of the experts are captured and presented to the students by the smart musical instrument in a way that requires minimal perceptual mapping. This would furthermore enable to take a lesson from a famous teacher that lives abroad.

## Conclusion

In this paper, we gave an overview of existing setups for augmented and smart musical instruments. Those setups aim to support students in learning the respective instrument, especially beginners.

From the presented setups, we derive recommendations for designing and implementing augmented and smart musical instruments that optimize the learning experience. Furthermore, smart and augmented musical instruments can enable music lessons with teachers over distance with reduced requirements on perceptual mappings.

While this paper focuses on keyboard and string instruments, the extension to other types of musical instruments such as wind instruments constitutes an integral part of future work.